\documentclass[thmsa]{article}
\usepackage{sw20lart}

\input{tcilatex}

\begin{document}

\author{Rafael A. Vera\thanks{%
rvera@udec.cl} \\
Departamento de F\'{\i}sica. \\
Facultad de Ciencias F\'{\i}sicas y Matem\'{a}ticas. \\
Universidad de Concepci\'{o}n. Chile}
\title{The new cosmological scenario and dark matter fixed by general
experimental facts }
\maketitle

\begin{abstract}
Both from gravitational (G) experiments and from a new theoretical approach
based on a particle model it is proved that the classical invariability of
the bodies, after a change of relative rest-position with respect to other
bodies, it is not true. The same holds for the traditional hypotheses based
on the classical one. The new relationships are strictly linear. From them
it is proved that a universe expansion must be associated with a G expansion
of every particle in it, in just the same proportion. It does not change the
relative distances, indefinitely. From the relative viewpoint, globally, the
universe must be rather static. According to the new cosmic scenario,
galaxies must be evolving, indefinitely, in rather closed cycles between
luminous and black states. The new kind of linear black hole must absorb
radiation until it can explode after releasing new H gas that would trigger
new luminous period of star clusters and galaxies. Statistically, most of
the galaxies must be in cool states. The last ones should account for all of
them, the higher velocities of the galaxies in clusters, the radiation
coming from intergalactic space, including the low temperature black-body
background observed in the CMBR.
\end{abstract}

\section{Introduction}

The current hypotheses in gravitation and \ in cosmology are tacitly based
on \textit{the classical hypothesis on the absolute invariability of the
bodies after a change of the rest position with respect to other bodies}.
Such hypothesis comes from the fact that, according to the Equivalence
Principle (EP), all of the bodies of his local system obey the same inertial
and gravitational (G) laws\cite{1}. Therefore, all of then must change in
the same proportion, i.e., every local ratio would remain constant.

In principle such changes can be tested by observers that have not changed
of position in the field. However this has not been done because, on the
contrary, \textit{current G tests are tacitly based on such classical
hypothesis, as shown below.}

\subsection{A crucial test for the cosmological hypotheses}

To fairly test such hypothesis the observer ($A$) must stay in a fixed
distance from the earth center. The observed body ($B$) must be at another
distance from it.

A well-defined experiment that meet this condition is the G time dilation
(GTD) experiments done with standard clocks, \cite{2}$^{,}$\cite{3}

From the results of such experiments, corrected after special relativity, it
is concluded that \textit{the relative frequency of a non local (NL) clock }$%
B$\textit{, located at rest at a well defined distance }$\Delta r$\textit{\
from earth surface, runs with a different frequency compared with the clock }%
$A$\textit{\ at the earth surface}. Thus \textit{the clocks are not
invariable after a change of distance with respect to the earth center}.

\subsubsection{Discussion}

It has been argued that: ''\textit{experiments to detect difference of
frequencies between the clock }$B$\textit{\ and }$A$\textit{\ have given
negative results. What happens (in a GTD experiment) is that an
electromagnetic signal controlled by clock B generally arrives at A with a
frequency different from the frequency of the same type of clock there.
Photons must do some work in moving around and their frequency changes}''.

This is not true because \textit{the readings of the clocks} \textit{do not
depend on the frequency or energy of any photon travelling between them.}

This fact is most obvious in the Hafele-Kerting experiments\cite{2} in which
no photon was used. The readins of the clocks $A$ and $B$ were directly
compared in the earth surface, before and after experiments. During
experiments of $48$ hours, the clocks $B$ were flying at $9$ km over the
earth surface. Thus the differences of the readings of the clocks did not
depend on the frequency of any photon travelling between such clocks%
\footnote{%
The same fact holds for other experiments in which the time interval between
the initial and final readings of the clock $B$ are obtained from
electromagnetic signals coming from such clock\cite{3}. Such time intervals
are differences of times in which the time of flight of the initial signals
is cancelled out by the one of the last signal. Thus such measurements do
not depend at all on the frequency of the photons.}.\newline
Then the differences of time intervals observed in the GTD experiments made
up with clocks can only be due to real differences of the relative
frequencies of the standard clock B, compared with A\footnote{%
An observer at $B$ cannot detect such change because, according to the
equivalence principle\cite{1}, all of the natural frequencies of the local
bodies have changed in the same proportion after a change of rest position
with respect to the earth center.} This means that , during the flight, some
fundamental physical change had occurred to every part of the clock $B$.

Then it is expected that \textit{the current relations between quantities
measured by observers located in different G potentials are not strictly
homogeneous because their reference clocks are not strictly the same with
respect to each other, respectively}. They would be sources of errors in the
current literature.\newline

It is also said, without fair demonstration, that the GTD experiments would
have verified the theory of GR. On the contrary, from them, the photons
emitted by the NL clock $B$ would \textquotedblleft start\textquotedblright\
their trips with an initial frequency shift $z$ with respect to the local
clock $A$. From the fact that the final redshift of the photons is just
equal to $z$ it is concluded that: \textquotedblleft \textit{during the trip 
}$BA$\textit{, the relative frequency of the photons, with respect to the
observer }$A$\textit{, remains constant}\textquotedblright .\ (Relative
frequency conservation law for photons.).

Then \textit{the G redshift of photons is not due to real frequency changes
of the free photons. It is due to differences of the natural frequencies of
atoms and clocks of observers located at rest in different distances from
the field source}. The same conclusion was obtained by Vera in1981 from
direct application of \textquotedblright wave continuity\textquotedblright 
\cite{4}.

\subsection*{ The non local form of the Equivalence Principle}

From above it is concluded that: to relate quantities measured in different
G potentials, they must be transformed to some well-defined reference
standard in a well-defined G position of the field. Here, the fixed position
of such observer (A) is stated by means of a subscript $a$.

From Lorenz equations and GTD experiments, it is inferred that the
``relative''\ quantities can depend on the velocity and distances of the
body and of the observer with respect to the field sources.(Vera 1981). On
the other hand, from the EP, when an observer moves altogether with his
clocks he finds that the local ratios between frequencies ($\nu $) masses ($%
m $) and lengths ($\lambda $) are constants that do not change after a
change of position of the measuring system with respect to the G field
sources. The opposite comes true when the observer $A$ remains in a fixed
position $a$. He finds, from GTD experiments, that the frequency of the
clock at rest at $B $ is a function on its position ($r$) in the field, say
\ $\nu _{a}(0,r)$. The first results can be consistent with the last ones
only if: ``\textit{the relative values of the frequencies, masses and
lengths of any well-defined part of the system have changed in just the same
proportion after the same change of relative position with respect to the G
field sources}''. Only in this way all of the local ratios can remain
unchanged. This may be called the NL form of the EP, or NL EP.

\section{The field equations fixed by experimental facts}

A short cut can be done from results of experiments and applications of the
NL EP. \newline

For example, assume that the observer A throws a clock upward with some
energy $\Delta E_{a}(0.r)$. From results of free fall experiments, the clock
would stop at some NL radius $r=a+\Delta r$ given by the three first members
of (1). From results of the GTD experiments made up by the observer $A$, the
clock at rest at $B$ would have a frequency $\nu _{a}(0,r)$ given by the
third and fourth member of (1)\footnote{%
The common unit of mass and energy used here is $1\left[ joule\right] $}.
The last member of (1) comes from the NL EP applied to any of the
frequencies, masses, lengths and wavelengths, of any particle or standing
wave of the same system.

\begin{equation}
\Delta \phi (r)=\frac{\Delta E_{a}(0,r)}{m_{a}(0,r)}=\frac{GM}{a}\frac{%
\Delta r}{r}=\frac{\Delta \nu _{a}(0,r)}{\nu _{a}(0,r)}=\frac{\Delta
m_{a}(0,r)}{m_{a}(0,r)}=\frac{\Delta \lambda _{a}(0,r)}{\lambda _{a}(0,r)}=%
\frac{1}{2}\frac{\Delta c_{a}(r)}{c_{a}(r)}  \tag{1}
\end{equation}

The last member results from the application of this equation to any
standing wave of the system. In it, $c_{a}(r)$ is the relative (NL) speed of
light at $B$ with respect to the observer $A$. This one is the product of
its relative frequency $\nu _{a}(0,r)$, and its relative wavelength $\lambda
_{a}(0,r)$.

In a previous article, in (1981), it is proved, step by step, that this
equation is consistent with the current tests for G theories\cite{4}.

\subsection{The relative mass-energy conservation \ }

From the 2$^{nd}$ and 5$^{th}$ members of (1), the relative mass of the
clock $B$, with respect to the observer $A$, depends on its relative
position:

\begin{equation}
m_{a}(0,r)=m_{a}(0,a)+\Delta E_{a}(0,r)  \tag{2}
\end{equation}

The energy $\Delta E_{a}(0,r)$ given up to the clock is not given up to the
G field: it remains ''in the clock'' as an additional mass. Vice versa,
during a free fall from $r$, its relative initial rest mass is $m_{a}(0,r)$.
From (2) and special relativity, its final mass passing by\ $A$\ is. 
\begin{equation}
m_{a}(V,a)=m_{a}(0,a)+\Delta E_{a}(0,r)=m_{a}(0,r)  \tag{3}
\end{equation}

\textit{During the free fall, the relative mass of the clock, with respect
to the observer A, remains constant}. This is consistent with the relative
frequency conservation law for photons derived above. Then it may be
concluded that\textit{\ there is not a true exchange of energy between
bodies or photons and the G field}. This result is in clear contradiction
with \textit{energy of the G field assumed by Einstein}\cite{4,5}.

\subsection{ The linear black hole}

The theoretical properties of the ``linear black hole'' (LBH) derived from
(1) \textit{are radically different from the conventional ones\cite{4} F}or $%
2GM>>r$, the gradient of the relative speed of light would produce
dielectric reflections preventing the escape of photons and nucleons\cite{4}%
. On the other hand it has a larger cross-section for photon capture. Thus
the average NL mass-energy of its nucleons increases with the time. When it
gets higher than the one in free state, the LBH can explode. Their neutrons
would decay into new hydrogen.

\subsection*{The theoretical field equation from the NL form of the EP}

From the NL EP, a \textit{particle model} made up or radiation in stationary
state between any two parts of a system \textit{must obey the same inertial
and gravitational laws of the particles in it}. This fact has been verified
by Vera (1981,1997) with a full consistency with special relativity, quantum
mechanics and equation (1) \cite{4,5}.\newline
When particle models emulate all of the uncharged particles of the universe
it is found that, according to the Huygen principle, the particles are the
result of constructive interference of wavelets crossing the space. \textit{%
The properties of the empty space in some position }$r^{i}$\textit{\ can
depend only on the actual perturbation frequency of the space produced by
all of the wavelets with random phases crossing it}. Each wavelet
contribution must be proportional to the product of its frequency and of its
amplitude. After taking into account the cosmological red shift, in which $%
d\nu /\nu $ \textit{= }$dr/R$, the average perturbation frequency of the
space, called $w(r^{i})$, must be proportional to:

\begin{equation}
w(r^{i})\propto \sum_{j=1}^{\infty }\frac{\nu ^{j}}{r^{ij}}\exp \left[ \frac{%
r^{ij}}{R}\right] =\sum_{j=1}^{\infty }\frac{m^{j}}{r^{ij}}\exp \left[ \frac{%
r^{ij}}{R}\right] \cong 4\pi \rho R^{2}  \tag{4}
\end{equation}

The \ average density of the universe, in $joules/m^{3}$ is $\rho $. The
Hubble radius is $R$.\newline

The best fit of (4) with (1) occurs for particles in equilibrium with the
space:

\begin{equation}
\lambda _{a}(0,r)w_{a}(r)=\text{ Constant}  \tag{5}
\end{equation}

\begin{equation}
\Delta \phi (r)=\frac{\Delta \lambda _{a}(0,r)}{\lambda _{a}(0,r)}=-\frac{%
\Delta w_{a}(r)}{w_{a}(r)}\text{ \ \ \ In which \ }G(r)=-\frac{1}{w_{a}(r)} 
\tag{6}
\end{equation}

From (6), the universe density is about 30 times the average density of
luminous matter. This is consistent with dark matter estimated in some 
\textit{clusters}.

\subsection{Matter expansion due to universe expansion}

Assume, as a hypothesis to be tested, that after a time \textit{dt} the
distances between the galaxies $i$ and $k$ have increased the proportion,

\begin{equation}
\frac{dr^{ik}}{r^{ik}}=Hdt  \tag{7}
\end{equation}

From (4), after using (6) and (7), it is found that the increase of the G
potential produces a gravitational expansion of any standard rod of length $%
\lambda $ given by:

\begin{equation}
d\phi (r)=\frac{d\lambda }{\lambda }=-\frac{dw}{w}=\frac{dr^{ik}}{r^{ik}}=%
\frac{dR}{R}=Hdt\text{ ;\ \ \ \ \ \ \ \ \ \ \ \ \ \ \ \ \ }\frac{\lambda }{R}%
=\text{ Constant\ \ }  \tag{8}
\end{equation}

\section{The new cosmological scenario fixed by experimental facts}

\textit{From equation }(8)\textit{\ it is concluded that: in the average,
the relative distances and the average density of the universe cannot change
with the time. }The universe age must be rather infinite.

\textit{The evolution of the celestial bodies must be occurring according to
rather closed cycles between luminous (hot) and non-luminous (cool) states. }

1) The hydrogen atoms, after stellar evolution, must be evolving,
indefinitely, in closed cycles between states of gas and linear black hole
(LBH), and vice versa.

2) The explosions of massive LBHs would fix the initial period of a luminous
star cluster or a galaxy. The new stars would be normally formed from
condensation of gas over older bodies that existed before the explosion.

3) A galaxy must also be running, almost indefinitely, in rather closed
cycles between luminous and dark states, and vice versa. Something similar
may hold for clusters.

Statistically, all of the evolution stages of the galaxies should be present
in the proportion fixed by their evolution periods. Since the
energy-recovering period of dark galaxies must by of a higher order of
magnitude that the luminous period of galaxies, then \textit{most of the
universe must be in the state of black galaxy cooled down by LBHs and the
rest of the universe}. They should account for most of the higher velocities
of galaxies in clusters and for the radiations coming from the intergalactic
space, mainly gamma, cosmic and low temperature blackbody radiation (CMB).

Galaxies should start their luminous period with new gas, free of heavy
metals, with a high density of randomly oriented angular momentum generated
during the LBH explosion. This should correspond with ``elliptical
galaxies''.

During the luminous period of a galaxy, the luminous volume would decrease
with the time. The last luminosity should come from a small region located
in its center, in the strong fields of massive bodies. They should
correspond with the true (radio noisy) \textit{quasars} of relatively
variable luminosity. Most of their red shifts would be gravitational one.
They should be not confused with QSOs of large Hubble red shift.

Most of the energy released in a matter cycle, from the state of gas up to
LBH, is gravitational. Most of it must be transformed into other kind of
energy around neutron stars. Thus the true role of the G energy in the
interpretation of the celestial phenomena is most important\cite{6}.

\section{Conclusions}

The classical hypothesis on the invariability of the bodies after a change
of relative position with respect to other bodies is not consistent with the
experimental facts. The true changes occurring to the bodies, after changes
of position in the G field must be described by using a position-dependent
formalism with respect to some well-defined observer that does not change of
position in the G field. In this way the fundamental errors derived from
this wrong hypothesis can be eliminated.

The new field relations derived from the EP and the G tests are strictly
linear ones. They rule out the presumed energy of the G field. The G field
does not exchange energy either with photons or with bodies. The G energy
comes from the bodies, not from the field.

The new approach based on particle models made up of photons in stationary
states provides more exact relationships for long range interactions and a
for unified understanding on the properties of bodies and their G fields.
The new relations reveal that, in the average, the relative density of the
university cannot change with the time. The average relative distances must
remain constant, indefinitely, i.e.,the universe age should be rather
infinite.

In the new scenario, the H gas must be evolving in rather closed cycles
between the states of gas and LBH and vice versa. A\ LBH, after absorbing
energy from the space, would explode. The new gas, condensed over other
bodies, would regenerate star clusters or galaxies. Galaxies would be
evolving, rather indefinitely, in rather closed cycles between hot and cool
states. Most of the matter of the universe must be in the black galaxies
that would be absorbing energy from the rest of the universe. They must
account for the black body radiation coming from the intergalactic space,
observed in the CMB. They must also account for the anomalous velocities of
galaxies in clusters.

The LBH explosions should account for the clean H and\ high densities of
angular momentum with random orientations observed in some galaxies. They
would be testimonies of the ''small bangs'' that occurred rather recently.

\end{document}